\documentclass[10pt]{iopart}
\usepackage{graphicx}

\usepackage{iopams}  
\usepackage{cite}
\usepackage{color}
\begin{document}

\title[Geometrical and Potential Time Delays in Gravitational Lensing]{An Examination of Geometrical and Potential\\ Time Delays in Gravitational Lensing}

\author{Oleg Yu. Tsupko$^1$, Gennady S. Bisnovatyi-Kogan$^{1,2,3}$, Adam Rogers$^4$ and Xinzhong Er$^5$}
\vspace{5pt}

\address{$^1$ Space Research Institute of Russian Academy of Sciences, Profsoyuznaya 84/32, Moscow 117997, Russia}
\address{$^2$ National Research Nuclear University MEPhI (Moscow Engineering Physics Institute), Kashirskoe Shosse 31, Moscow 115409, Russia}
\address{$^3$ Moscow Institute of Physics and Technology, 9 Institutskiy per., Dolgoprudny, Moscow Region, 141701, Russia}
\address{$^4$ Department of Physics and Astronomy, University of Manitoba, Winnipeg, MB, R3T 2N2, Canada }
\address{$^5$ South-Western Institute for Astronomy Research, Yunnan University, Kunming 650000, China}

\eads{\mailto{tsupko@iki.rssi.ru}, \mailto{tsupkooleg@gmail.com}, \mailto{gkogan@iki.rssi.ru}, \mailto{rogers@physics.umanitoba.ca}, \mailto{xer@ynu.edu.cn}}
\vspace{10pt}

\begin{indented}
\item[]ORCID numbers:
\item[]Oleg Tsupko: 0000-0002-2159-8350;
\item[]Gennady Bisnovatyi-Kogan: 0000-0002-2981-664X;
\item[]Adam Rogers: 0000-0003-2953-2054;
\item[]Xinzhong Er: 0000-0002-8700-3671
\end{indented}

\vspace{10pt}

\begin{indented}
\item[]April 2020
\end{indented}

\begin{abstract}
In this paper we investigate the relation between the potential and geometric time delays in gravitational lensing. In the original paper of Shapiro (1964), it is stated that there is a time delay
in the radar signals between Earth and Venus that pass near a massive object (the Sun), compared to the path taken in the absence of any mass. The reason for this delay is connected with the influence of gravity on the coordinate velocity of a light ray in a gravitational potential. The contribution from the change of the path length, which happens to be of second order, is considered as negligible. Nevertheless, in the gravitational lens theory the geometrical delay, related to the change of path length, is routinely taken into account along with the potential term. In this work we explain this apparent discrepancy. We address the contribution of the geometric part of the time delay in different situations, and introduce a unified treatment with two limiting regimes of lensing. One of these limits corresponds to the time delay experiments near the Sun where the geometrical delay is shown to be negligible. The second corresponds to the typical gravitational lens scenario with multiple imaging where the geometrical delay is shown to be significant. We introduce a compact, analytical, and quantitative criteria based on relation between the angular position of source and the Einstein radius. This criterion allows one to find out easily when it is necessary to take the geometrical delay into account. In particular, it is shown that the geometrical delay is non-negligible in the case of good alignment between source, lens and observer, because in such a case it becomes a first order quantity (the same order as the potential term).
\end{abstract}

%
\vspace{2pc}

\section{Introduction}

The time delay of a lensed light ray, relative to an undeflected ray, is the amount by which light is delayed during propagation in the presence of a gravitating body (a lens) in comparison with straight line propagation to an observer in flat spacetime. In the original paper of Shapiro \cite{Shapiro-1964} it is stated that the effect of delay is
related to the decrease of the coordinate velocity of light of a ray during propagation through a region of gravitational potential and that the contribution to time delay from the change in path, being of second order, is negligible. Nevertheless, in gravitational lens theory the geometrical delay, related to the change of path length, is routinely taken into account along with the potential  term \cite{GL2, Cooke-1975}.

\begin{figure*}
\begin{center}
\includegraphics[width=0.95\textwidth]{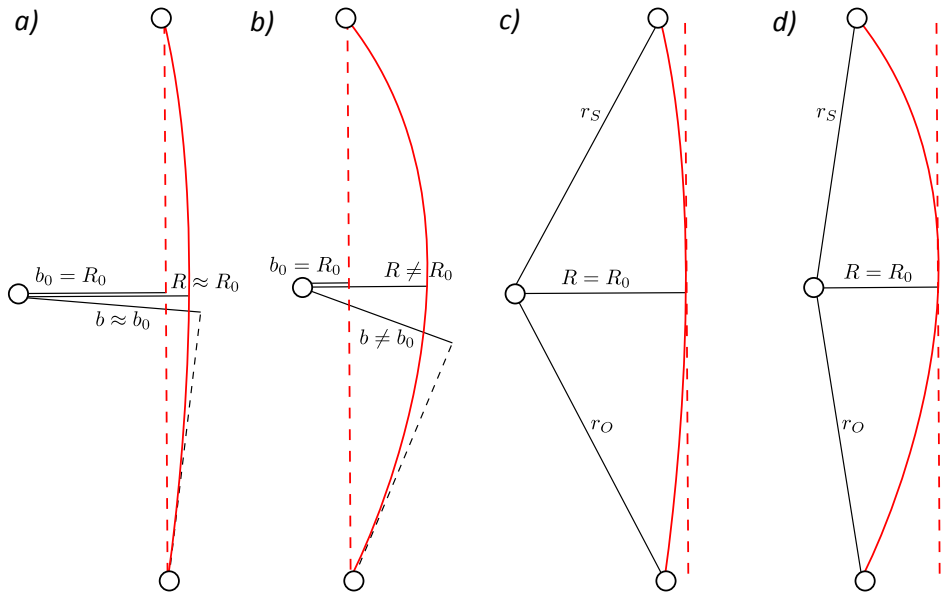}
\end{center}
\caption{In this figure we illustrate the discrepancy between the trajectory of an actual ray (curved solid line) and an undeflected ray (straight dashed line). The rays have impact parameters $b$ (actual) and $b_0$ (undeflected) and distance of closest approach to the lens $R$ and $R_0$ respectively. Panels a and b show the unperturbed path from source to observer and panels c and d  represent unperturbed paths with the same distance of closest approach. In all cases the deflection of the ray paths appear greatly exaggerated. The distance of closest approach is assumed to be much larger than the Schwarzschild radius and we may safely assume small angles. This picture shows that in case of good alignment of source, lens and observer (cases b and d), the difference between real and unperturbed trajectories
becomes significant, leading to restrictions on validity of approximations used in derivation of Shapiro's delay. Panels c and d looks similar, but in case c the unperturbed ray is close to the positions of the source and observer, while in case d the unperturbed ray is very far from the positions of the source and observer. Quantitative difference between cases c and d leads to qualitative difference: in case d, geometrical delay related to the change of path length becomes  a first order quantity (the same order as the potential term) and cannot be neglected. See also Fig.3 where two regimes are shown. See the discussion in Sec.\ref{sec:shap-deriv}. }
\label{fig:shap}
\end{figure*}

Surprisingly, it seems that there is no explicit explanation of this apparent discrepancy in the current literature. In this paper, we address this question and examine when the geometric part of the time delay can or cannot be neglected, based on the following ideas:

(i) When deriving the Shapiro delay, it is assumed that the trajectory of a light ray in the presence of a gravitating body does not change significantly. That is, the impact parameter of the deflected ray is considered as {\it almost equal} to the impact parameter of the unlensed straight line path (see Fig. \ref{fig:shap}a). However, this approximation is not justified if the angular position of the source is close to the observer-lens line. In this case, the impact parameter of the lensed ray is substantially different than the unlensed ray (see Fig. \ref{fig:shap}b), and the trajectory is changed significantly. This leads us to the idea that in such a case it is necessary to take into account the change of path length, which is the geometric delay.

\begin{figure}
\begin{center}
\includegraphics[width=0.55\textwidth]{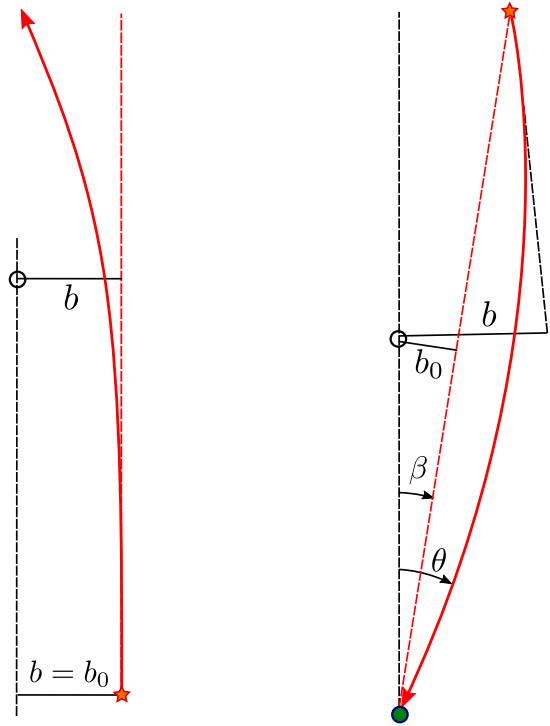}
\end{center}
\caption{Difference between impact parameter $b$ of lensed ray (solid curve) and impact parameter $b_0$ of unlensed ray (dashed straight line) in gravitational lensing. Left: configuration, when lensed light ray and the imaginary undeflected ray are emitted by the same source in the same initial direction, so they have the same impact parameters, $b=b_0$. After deflection the actual light ray goes to another point compared with an undeflected ray. Right: formation of primary image during gravitational lensing. Opposite to left panel, the positions of source and observer are fixed, and the deflected light ray has the same origin and end point as an undeflected ray. Therefore the deflected light ray should have another impact parameter $b$ than the impact parameter $b_0$ of unlensed light ray, because it is emitted in different initial direction. This holds for both primary and secondary images. Due to the difference between $b$ and $b_0$, the image position $\theta$ seen by the observer is different from the source position $\beta$.}
\label{fig:imp-par}
\end{figure}

(ii) On the other hand, in the usual approach of gravitational lensing (e.g., when a change in the apparent position of the source or multiple images are considered), the impact parameters of the lensed and unlensed rays are treated as two {\it non-equal} quantities from the very beginning (see Fig.~\ref{fig:imp-par}). This difference is responsible for changing the angular position of the image compared to the angular position of the source.

In fact, the difference between the impact parameters of the lensed and unlensed rays can be small or large, depending on the relative position of the objects. The difference is the largest when the source, lens and observer are perfectly aligned, because in this case the impact parameter of unlensed ray goes to zero (see Fig. \ref{fig:shap}b). In this scenario the approximations used in deriving the Shapiro delay are obviously not satisfied and the change of the path cannot be neglected, thus the geometrical delay needs to be taken into account. In particular, the geometrical delay becomes of crucial importance in the typical gravitational lensing scenario, when the source and lens are well-aligned and multiple images are formed.

In order to quantitatively describe when it is possible to neglect the geometric delay, we start by analyzing the difference between the impact parameters of the lensed and unlensed rays for the point-mass gravitational lens. Following standard notation in the gravitational lens literature, we denote the angular positions of images and the source as $\theta$ and $\beta$, respectively. The angular size of the Einstein ring, or Einstein radius, is $\theta_E$. We consider the primary and secondary images, and show that the difference is negligible only for the primary image in the case of $\beta \gg \theta_E$.

\begin{figure*}
\begin{center}
\includegraphics[width=0.95\textwidth]{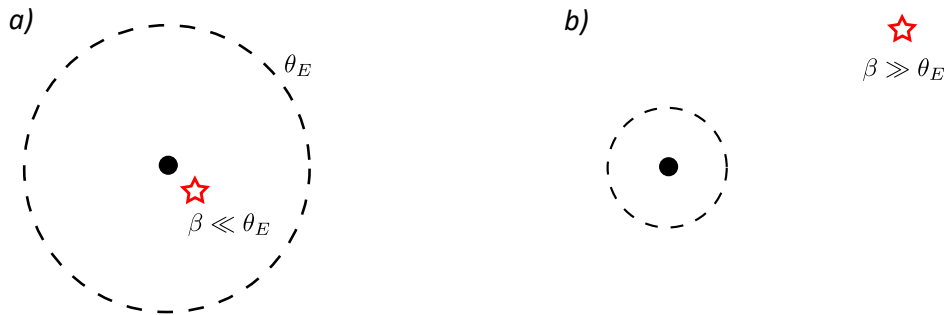}
\end{center}
\caption{Two regimes of lensing. The lens (black spot), the Einstein radius (dashed black circle) and the source are shown. a) Nearly perfect alignment of the source, the lens and the observer, $\beta \ll \theta_E$. We show that in this regime, the geometrical delay for the primary image is significant (first order). b) Large angular separation between the source and the lens, $\beta \gg \theta_E$. In this regime, the geometrical delay for primary image is small (second order). Regarding secondary images, the geometrical delay should be taken into account in both regimes.}
\label{fig:two-regimes}
\end{figure*}

Based on this analysis, we introduce two limiting regimes of lensing. In the first regime, the source, lens and observer are nearly perfectly aligned, i.e. the angular separation between the source and the lens is much smaller than the Einstein radius, $\beta \ll \theta_E$, see Fig.~\ref{fig:two-regimes}a. In this case, the difference between the impact parameters of the lensed and unlensed rays is considerable. We show that in this lensing regime the potential and the geometrical terms are comparable. Therefore, it is not possible to neglect the geometrical term in time delay calculations.

In the second regime, the angular separation between the lens and source is relatively large, i.e. $\beta \gg \theta_E$, see Fig.~\ref{fig:two-regimes}b. Then the change of trajectory of light from the primary image is not significant, and the impact parameter of the lensed trajectory can be considered equal to the unlensed one. We show that in this case, the geometrical term can indeed be neglected in comparison with the potential term and Shapiro's formula is recovered.

To summarize, here we show that the geometrical delay related to the change of path length is of the first order in one limiting case and is of the second order in another limiting case. This explains the apparent discrepancy discussed here. In experiments with light propagation near the Sun we deal with the primary image only and the regime $\beta \gg \theta_E$. Therefore, the geometrical delay can be neglected. In gravitational lens systems with multiple images, we often deal with a good alignment of source, lens and observer, and image positions $\theta \simeq \theta_E$. This roughly corresponds to our second limiting regime $\beta \ll \theta_E$ (or, at least, with the case $\beta \lesssim \theta_E$). Therefore the geometrical delay should be taken into account for every image.

This paper is organized as follows. In the next section we present a short overview of selected literature about the lensing time delay. In Sec.\ref{sec:lens-overview} we review the main properties of the point-mass lens equation and its solutions. In Sec.\ref{sec:shap-deriv} we discuss the usual ways of deriving the time delay and demonstrate peculiarities that occur in the case of good alignment of source, lens and observer. In Sec.~\ref{sec:b-b0} the analysis of difference between the impact parameters of the lensed and unlensed rays for a point-mass gravitational lens is presented. In Sec.~\ref{sec:nearly-perfect} and \ref{sec:large-ang-sep} we investigate the time delay in case of nearly perfect alignment, and large separation of source and lens respectively. Sec.~\ref{sec:summary} is the summary. In Appendix we present additional comments about the deflection angle in two limiting regimes of lensing.

\section{Some historical notes}

After the successful measurement of light deflection during a solar eclipse (Eddington, 1919), Shapiro suggested a measurement of time delay in the Solar system as a fourth test of General Relativity \cite{Shapiro-1964}. In Shapiro's proposal, one can send intense radar pulses towards Venus or Mercury and measure the round-trip arrival time of echoes. To see the effect of time delay due to the gravitational influence of the Sun, a radar beam should be sent towards the planet when it is nearly opposite to the Earth on the far side of the Sun, and the path of the pulse grazes the Solar limb. The physical explanation of the Shapiro delay is in the change of the speed of light in the presence of a potential well of a gravitating body (in this case, the Sun). Subsequently, the Shapiro delay in the Solar system has been successfully measured (see, for example, \cite{Shapiro-1968, Shapiro-1971}).

At the same time, parallel to Shapiro's investigations, scientists began to discuss the time delay effect in gravitational lens systems with multiple images. Refsdal \cite{Refsdal-1964a, Refsdal-1964b} studied the full properties of the point-mass lens (see also works of Klimov \cite{Klimov-1963}, Liebes \cite{Liebes-1964} and Byalko \cite{Byalko-1970}), and calculated the difference of the light travel times between the two lensed images of a source. It was shown that the Hubble constant and the mass of the lens can be determined from a measurement of the time delay between different images and image positions. The potential application of the time delay between images of a distant supernova (SN) seen through a distant galaxy close to the line of sight was also discussed.

Subsequently, the theory of time delay in gravitational lensing has been extensively studied. Cooke and Kantowski \cite{Cooke-1975} argued that the time delay consists of two terms: the first is due to the difference in geometrical path lengths (the geometrical delay), and the second is due to the difference in the gravitational potential through which the light ray travels (the potential delay). The total time delay was conveniently written in terms of the Fermat potential in \cite{Schneider-1985}. Fermat’s principle and the time delay surface were studied by Blandford and Narayan \cite{Blandford-1986}.

Nowadays, the idea of Refsdal is applied to cosmological studies using strong lens systems. In cases where the background source intensity varies, such as an active galactic nucleus (AGN) or a SN, the variability pattern in each of the multiple images is delayed in time due to the independent light path length and the change in gravitational potential along these paths \cite{Shapiro-1964,Refsdal-1964b,GL1, Suyu-2010}.

A thorough understanding of gravitational lens time delay is of vital importance to a number of topics on the cutting edge of modern astrophysics. For example, time delay effects are a valuable tool for testing the validity of General Relativity using binary pulsars \cite{BK-pulsars}. Time delay tests will become more important as exotic binary pairs, such as pulsar-black hole binaries that contain more mass that binary pulsars, are discovered in future studies \cite{PSR}. On larger scales, images of the first lensed supernova, SN Refsdal, were first detected in 2015 \cite{SNE1}. This core-collapse supernova produced multiple, resolved images that were observed through the foreground galaxy cluster lens. Predictions for the appearance of further images were made based on the time delays along individual image paths \cite{SNE2} and the images subsequently observed \cite{SNE3}. So far only one other lensed supernova has been discovered \cite{SNE4}, however more of these systems are expected to be found in the future \cite{SNE5}. Currently, the locally measured value of the Hubble constant differs at high confidence from its value estimated using the Planck Cosmic Microwave Background (CMB) data, a situation which is known as the Hubble tension \cite{tension, planck2018-01, planck2018-06, h0-time-delay}. Different explanations of the Hubble tension have been proposed \cite{tens-solv-01,tens-solv-02,tens-solv-03, tens-solv-04,tens-solv-05, Handley-2019,tens-solv-bk}. To resolve this apparent paradox, cosmological methods independent from CMB, BAO and SNe are necessary, e.g. \cite{tips, ruler}. Strong lensing time delay measurements offer such a possibility \cite{Refsdal-1964b}. Lensing has little sensitivity to other cosmological parameters and thus can minimize the degeneracy between different cosmological models (e.g. \cite{suyu2018, h0-time-delay}).

\section{Point-mass lens equation in gravitational lensing}
\label{sec:lens-overview}

Here we briefly present well known properties of the point-mass lens equation.

\begin{figure}[h]
	\begin{center}
		\includegraphics[width=0.9\textwidth]{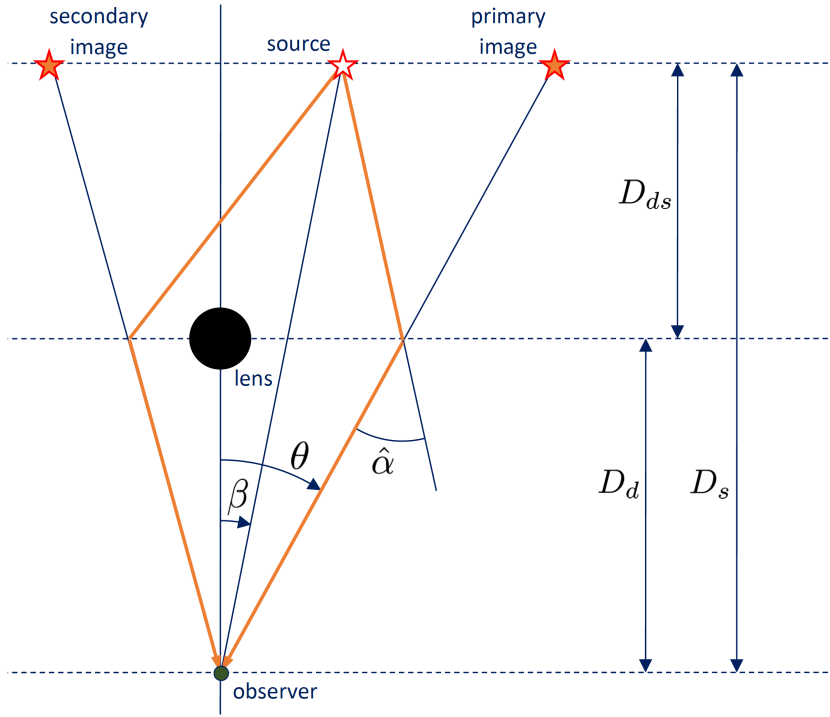}
	\end{center}
	\caption{Point-mass-lens with two images}
	\label{fig:geometryl}
\end{figure}

The lens equation relates the position $\beta$ of a source and positions $\theta$ of the corresponding images through the deflection angle $\hat{\alpha}$. For the typical geometry to describe lensing and our adopted notation, see Fig. \ref{fig:geometryl}.
In general, the small angle approximation and the thin lens approximation are adopted, i.e.
$\beta, \theta, \hat{\alpha} \ll 1$, and the spatial extension of the lens is much smaller than the distance between the source and the lens $D_{ds}$, or the distance between the lens and the observer $D_d$.
For mathematical simplicity we assume an axially symmetric lens mass distribution. Thus, the lens equation has a simple form e.g.\cite{GL1,Keeton-2018}
\begin{equation} \label{le}
\beta = \theta - \frac{D_{ds}}{D_s} \hat{\alpha} \, ,
\end{equation}
where $\hat{\alpha}$ is the photon deflection angle. By the usual convention, $\beta$ is positive, and $\theta$ can be both positive (image is on the same side from lens as a source) and negative (image is on the opposite side). The case $\beta=0$ corresponds to perfect alignment of source, lens and observer. For a more general lens equations we refer to \cite{Frittelli-2000, Perlick-2004b, Virbhadra-2000, Bozza-Sereno-2006, Bozza-2008b, Aazami-2011a, Aazami-2011b}.

For a Schwarzschild point-mass lens the deflection angle is
proportional to the Schwarzschild radius $R_S$
\begin{equation} \label{einst-angle}
\hat{\alpha} = \frac{2 R_S}{b} \, , \quad R_S = \frac{2GM}{c^2} \, ,
\end{equation}
where $b$ is the lensed photon impact parameter and $M$ is the black hole mass. Substituting angle (\ref{einst-angle}) into (\ref{le}) and using $b = D_d \theta$, we recover the well-known lens equation for the point-mass lens:
\begin{equation} \label{le-point}
\beta = \theta - \frac{\theta_E^2}{\theta}  \,  .
\end{equation}
Here $\theta_E$ is given by
\begin{equation} \label{einst-radius}
\theta_E = \sqrt{ \frac{4GM D_{ds}}{c^2 D_d D_s} }  \,  ,
\end{equation}
which corresponds to the size of the Einstein ring observed in the case of perfect alignment. The lens equation (\ref{le-point}) has two solutions,
\begin{equation} \label{sols-point}
\theta_\pm = \frac{\beta}{2} \pm \frac{\beta}{2}  \sqrt{1+\frac{4\theta_E^2}{\beta^2} }    \,  .
\end{equation}
Usually $\theta_+ > 0$ is called the primary image (it is located at the same side from lens as a source), and $\theta_- < 0$ is called the secondary on the opposite side of lens in comparison with a source, see Fig.~\ref{fig:geometryl}.

The Einstein deflection angle (\ref{einst-angle}) uses the impact parameter $b$. Working in the small deflection approximation, this formula can be written using the distance of closest approach $R$ (the minimum value of the $r$-coordinate at the trajectory): $\hat{\alpha}=2R_S/R$. In textbooks, the written form of the deflection angle is usually determined based on the method used to derive it. This can be shown by expanding the exact integral expression for the deflection angle in powers of $M/b$ and $M/R$. The coefficients in these expansions will coincide to first order (e.g., \cite{Keeton-2005}). The difference only becomes apparent at second order. Since high order terms are not taken into account in the weak deflection approximation, in gravitational lensing the concepts of impact parameter and the closest approach distance are usually not distinguished (e.g., p.25 in \cite{GL1}).
In this article, in all first-order formulas (in particular, in all formulas in the next section), these quantities can be replaced with each other without loss of accuracy.

\section{Analysis of different derivations of the time delay and approximations used}
\label{sec:shap-deriv}

In this section, we review two approaches to the derivation of the Shapiro delay in the literature. We show that both of them cannot be applied to the situation where the source is very close to the observer-lens line, because the change of path becomes significant. Next, we review the derivation of the total delay (consisting of contributions from both the potential and geometrical terms) commonly used in gravitational lensing.

To begin, let us clarify the terminology used in this paper. The `time delay' $\Delta t$ is used for the time delay relative to the undeflected straight line ray. It means that $\Delta t= t -t_0$, where $t$ is the time travel in presence of gravitating body (lens) and $t_0$ is the travel time along an undeflected path in flat spacetime. However, the undeflected travel time $t_0$ is unknown in the case of strong lensing due to the slow relative motion between source and lens. In practice, it is impossible to actually measure such a time delay. The time delay in strong lensing generally refers to the `time delay between multiple images'. We will denote this delay as $\Delta = t_2-t_1$, and it is often referred to simply as 'time delay' in the strong lensing community. For example, in pp.126-127 of Schneider, Ehlers and Falco \cite{GL1}, there is a `time delay of a kinematically possible ray relative to the undeflected ray' ($\Delta t$ in our notations) and `arrival time difference (or time delay) for two images' ($\Delta$). In Section 3.4.2 of Congdon and Keeton's book \cite{Keeton-2018}, `time delay' and `differential time delay' are introduced. Obviously, the time delay between two images can be presented as a difference between time delays for every image, i.e. $\Delta = t_2 - t_1 = (t_2 - t_0) - (t_1 - t_0) = \Delta t_2 - \Delta t_1$.

In this paper, we use the terminology of gravitational lensing. In the discussion of the Shapiro experiment, the planet which reflects the signal is considered as the 'source' and the Sun is the 'lens'. In all formulas, the one-way travel delay is calculated.

\subsection{Derivation of Shapiro delay, first method}

In the first approach, one can integrate the gravitational potential along an unperturbed straight path, which results in a travel time containing the first-order $GM$-scale correction. This approach can be found at pp. 1106-1107 in Misner, Thorne and Wheeler \cite{MTW}, pp. 125-127 in Ohanian \cite{Ohanian-1976}, p.237 in Rindler \cite{Rindler-2006}, pp.25-26 in Dodelson \cite{Dodelson-2018}. The time of one-way travel from source to observer in terms of the coordinate time is:
\begin{equation} \label{travel-time-1}
t = \frac{D_{ds}}{c} \,  + \,  \frac{D_d}{c} \, + \, \frac{2GM}{c^3} \left[ \ln \frac{D_d + \sqrt{D_d^2 + b_0^2}}{b_0} \right. +
\end{equation}
$$
+ \, \ln \left. \frac{D_{ds} + \sqrt{D_{ds}^2 + b_0^2}}{b_0} \right] \, ,
$$
where the logarithmic terms are the Shapiro delay. Note that more complicated version of the formula is presented in Shapiro's paper \cite{Shapiro-1964}, because proper time at the Earth is used there, see also p.128 in \cite{Ohanian-1976} and p.1107 in \cite{MTW}.
The detailed derivation of formula (1) in Shapiro \cite{Shapiro-1964} can be found, for example, in \cite{mathpages}.

Usually we have $b_0 \ll D$ in practical situations, thus the formula can be simplified to \cite{Ohanian-1976, Rindler-2006, Dodelson-2018}
\begin{equation} \label{Shap-b}
\Delta t_{\textrm{Shap}} =  \frac{2GM}{c^3} \ln  \frac{4 D_d D_{ds}}{b_0^2}  \, .
\end{equation}

In such way of derivation, the gravitational bending of the ray is ignored in the computation \cite{MTW, Ohanian-1976}.
The indicated reason is that the difference in length (as well as the corresponding time delays) between the straight and curved paths
are proportional to the squared deflection angle. Since $\hat{\alpha}^2 \propto (GM)^2$, the possible correction due to bending is of the second order in $GM$ \cite{MTW, Ohanian-1976}. Since bending of path is ignored, the formula (\ref{Shap-b}) uses the impact parameter $b_0$ of unperturbed path. In other words, the difference between the impact parameter $b$ of the real trajectory and the impact parameter of an unperturbed light ray $b_0$ is considered as negligible.

Here we note that this approximation is valid as long as the source is not well aligned with the observer and the lens (Fig.\ref{fig:shap}a). If the angular separation between the lens and source becomes small (Fig.\ref{fig:shap}b), the impact parameter $b$ differs substantially from $b_0$. It means that the geometrical change of the path becomes significant. The same argument can be addressed for the distance of the closest approach $R$ and $R_0$.

In case of perfect alignment, the unperturbed impact parameter goes to zero, $b_0 \to 0$, and deflection angle is proportional to the Einstein angular radius, $\hat{\alpha} \propto \theta_E \propto \sqrt{GM}$.
Therefore, the correction due to the path bending becomes a first-order effect and cannot be neglected, i.e. $\hat{\alpha}^2 \propto \theta_E^2 \propto GM$. The length of the curved trajectory can be approximated by two straight lines $\sqrt{D_d^2+b^2} + \sqrt{D_{ds}^2+b^2}$. In the perfect alignment, $b \propto \sqrt{GM}$, the length of the curved trajectory differs from the length $D_d+D_{ds}$ of straight path by terms of the $GM$-order.

\subsection{Derivation of Shapiro delay, second method}

In the second approach, one can use the curved path and expansion of the exact integral for the travel time. This approach can be found at pp. 201-203 in Weinberg \cite{Weinberg-1972}, pp. 146-147 in Wald \cite{Wald-1984}, pp.236-238 in Hobson, Efstathiou, and Lasenby \cite{Hobson}, pp.173-176 in Bambi \cite{Bambi-2018}, and Perlick \cite{Perlick-lectures}. Although the integration is performed along the curved path, the calculation only shows retardation of time due to the gravitational potential and does not take into account the difference between the length of real and unperturbed paths (geometrical delay).

The exact integral for the travel time can be written by a source position $r_S$, observer position $r_O$ and the distance of the closest approach $R$ of curved ray (see also Fig.\ref{fig:shap}c,d). Then the integrand is expanded to the first order in $GM$ and integrated, giving the following expression for the one-way travel time:
\begin{equation}
t = \frac{(r_S^2- R^2)^{1/2}}{c} + \frac{(r_O^2- R^2)^{1/2}}{c} +
\end{equation}
$$
 + \frac{2GM}{c^3} \ln \frac{r_O + (r_O^2- R^2)^{1/2}}{R}  + \frac{GM}{c^3} \left( \frac{r_O - R}{r_O + R} \right)^{1/2}  +
$$
$$
+ \frac{2GM}{c^3} \ln \frac{r_S + (r_S^2- R^2)^{1/2}}{R} + \frac{GM}{c^3} \left( \frac{r_S - R}{r_S + R} \right)^{1/2}  \,  .
$$
Null terms are interpreted as travel time in absence of lens and linear terms are interpreted as Shapiro delay.
In practical situations we have $R \ll r_O,r_S$, so Shapiro (one-way) delay takes form \cite{Weinberg-1972, Hobson, Bambi-2018}:
\begin{equation} \label{shap-R}
\Delta t_{\textrm{Shap}}  = \frac{2GM}{c^3} \left( \ln \frac{4 r_S r_O}{R^2} + 1 \right) \, .
\end{equation}

Since integration is performed along the curved path, the second approach is usually argued as being more accurate (e.g., see discussion in \cite{Ashby-2008, mathpages}), which ends up with an additional term in eq.(\ref{shap-R}) in comparison with eq.(\ref{Shap-b}).
In the papers \cite{Rafikov-2005} and \cite{Hackmann-2019}, authors also argue that the Shapiro delay should be calculated along the lensing path. Additionally, it is possible to derive higher-order terms in the Shapiro delay connected with high-order terms in the metric, see, e.g., papers \cite{second-order-1983, Dymnikova-1984, Dymnikova-1986, second-order-01, second-order-02, second-order-turyshev}. In the paper \cite{Hackmann-2019}, the authors compare the exact numerical calculation of the time delay with different approximate formulas, including second-order Shapiro terms and the geometrical delay.

In this derivation method, a similar problem appears in the case of a well-aligned source and lens, see Fig. \ref{fig:shap}c,d. The null term contains the length of unperturbed path which is calculated as $(r_S^2- R^2)^{1/2} + (r_O^2- R^2)^{1/2}$. It means that the distance of the closest approach $R_0$ of an unperturbed light ray is assumed to be equal to $R$ of the real curved ray. But the imaginary straight line light ray having $R_0=R$ does not connect the source and the observer, as can be easily seen in Fig. \ref{fig:shap}c,d. Therefore, this approximation works well only if the line connecting observer and source is far from the lens (Fig. \ref{fig:shap}c). If the line connecting the observer and the source is close to the lens, the unperturbed ray is very far from the positions of the source and observer (Fig. \ref{fig:shap}d).

To conclude these two subsections, both derivations of the Shapiro delay used in the literature encounter difficulties when the angular separation between the source and lens is relatively small (good alignment). In short, this is due to the fact that the approximation $b \approx b_0$ (or $R \approx R_0$) breaks down. Therefore, it is not possible to calculate the first-order time delay without taking the change of path length into account.

\subsection{Derivation of time delay in gravitational lensing}

In contrast to the particular method of deriving the Shapiro delay, the gravitational lensing literature usually considers the impact parameter of an lensed ray different from the impact parameter of an unperturbed straight path. Consequently, the derivation of the time delay is different, since the change in path is taken into account, and the geometrical delay appears.

The usual approach to derive the lensing time delay is through the summation of two separate delays, namely the potential (Shapiro) and geometric delays. For example, on pp.125-127 of \cite{GL1}, we find that the travel time is calculated as
\begin{equation}
t = c^{-1} \int \left( 1 - \frac{2U}{c^2} \right) dl  =   c^{-1} l - 2 c^{-3} \int U dl \, ,
\end{equation}
where $l$ is the Euclidean length of the lensed path consisting of two straight lines with a bend near the lens, and the potential term is to be integrated along that path ($U<0$ is the Newtonian potential). Further, the subtraction of the travel time for an unlensed ray gives the time delay with two terms. The same ideas are presented in \cite{Cooke-1975}. \footnote{It is stated in \cite{Cooke-1975} that the photon travel time consists of two terms: '{\it The first term gives the time due to the length of path traveled and must be computed to first order in $G$. The second
is due to the potential well through which the photon traveled and is already first order in $G$. The difference in travel
times for two photons depends on the difference in both terms. For small scattering angles and point-mass deflectors
the two time delays are of the same order}'. See also p.53 of Schneider, Kochanek and Wambsganss \cite{GL2}. }

In this approach to the derivation, two things are taken into account: the curvature of the trajectory during the integration of the potential term (as in the previous subsection) and the change of path length which is responsible for the geometrical delay.

Note that, since the potential term is integrated along the lensed path, the resulting potential term is different from the Shapiro formula (\ref{Shap-b}). During integration, the actual impact parameter $b$ but not $b_0$ is used. Therefore, the final potential term is a function of image position $\theta = b/D_d$ but not source position $\beta=b_0/D_d$.

As we see, in the derivation used in gravitational lensing we take the geometrical delay into account. Our main goal in this paper is to understand why we do that in gravitational lensing and don't do this in the derivation of Shapiro's delay.

Apparently, it would be very helpful to derive the same two terms of the same order from the expansion of the exact integral expression for the time delay. As we discussed above, in the case of good alignment, the additional terms of the first order arise due to the significant change of the path length. Therefore, in order to derive the time delay expression for this case, we can assume good alignment from very beginning, e.g. $\theta \sim \theta_E$ (in the next Sections we will formulate the quantitative condition of good alignment). Such a rigorous derivation has been presented recently in the book of Congdon and Keeton \cite{Keeton-2018}, see also paper \cite{Keeton-2005} of Keeton and Petters.

In \cite{Keeton-2018}, the time delay analysis involves three scales: the mass parameter $m=GM/c^2$, the distance of closest approach to the deflector $R_0$ and the large radius $R_D$ characterizing the scale of the source and observer distances (all the distances $D_d$, $D_s$, $D_{ds}$ are simplified as $R_D$ assuming that they are of the same order). In the Taylor expansion for the arrival time of the lensed path, there are two arguments: $m/R_0$ and $R_0/R_D$. Further, authors \cite{Keeton-2018} assume that the distance of the closest approach is close to linear Einstein radius, $R_0 \sim R_E$, where $R_E = D_d \theta_E$.
Using $R_0 \sim R_E \sim \sqrt{mR_D}$, one can obtain
\begin{equation}
\frac{m}{R_0} \sim \frac{R_0}{R_D} \sim \left( \frac{m}{R_D} \right)^{1/2}.
\end{equation}
Therefore, during the Taylor expansion procedure, it is necessary to take into account both of the terms $m/R_0$ and $R_0/R_D$. Finally, this leads to the expression (3.110) of \cite{Keeton-2018} for the time delay in its usual (for gravitational lensing) form, with both potential and geometrical components:
\begin{equation}
\Delta t = \frac{D_d D_s}{2 c D_{ds}} ( \theta - \beta )^2  - \frac{4GM}{c^3} \ln {|\theta| }    \, .
\end{equation}

\section{Impact parameters of lensed and unlensed light rays}
\label{sec:b-b0}

Let us introduce a quantitative criteria for source, lens and observer alignment, when the change of path must necessarily be included. We will use this quantity to compare the behaviour of the impact parameters of lensed and unlensed light rays. In this section we discuss how the difference $(b-b_0)$ behaves depending on $\beta$. Working in the small angle approximation, we can write
\begin{equation} \label{b-b0}
b-b_0 = D_d (\theta - \beta) \, .
\end{equation}

\begin{figure}[h]
	\begin{center}
		\includegraphics[width=0.95\textwidth]{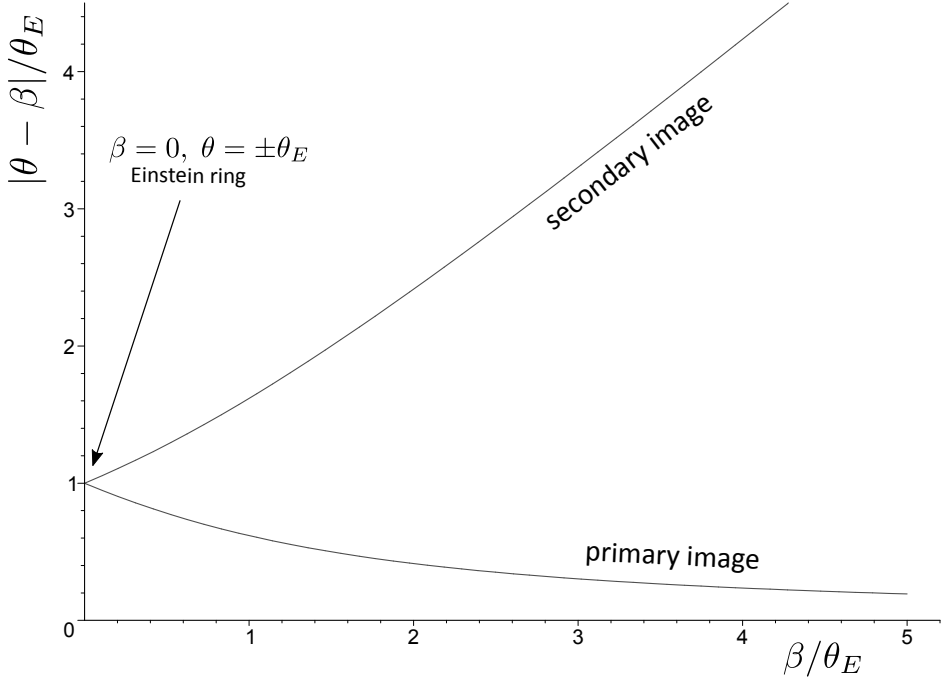}
	\end{center}
	\caption{The difference $|\theta - \beta|/\theta_E$ for primary and secondary images as a function of $\beta/\theta_E$. The difference is small only for primary image and large values of $\beta/\theta_E$. Using formula (\ref{b-b0}), we can say that the difference $(b-b_0)$ between impact parameters of lensed and unlensed rays is negligible only in this situation.}
	\label{fig:Delta}
\end{figure}

In Fig.\ref{fig:Delta} we plot $|\theta - \beta|/\theta_E$ as a function of source position for primary ($\theta \to \theta_+$) and secondary images ($\theta \to \theta_-$), where $\theta_\pm$ are defined in eq.(\ref{sols-point}). We conclude that the difference between impact parameters of lensed and unlensed rays is negligible only for the primary image when $\beta \gg \theta_E$.

Based on that, we introduce two limiting regimes of lensing: $\beta \ll \theta_E$ (nearly perfect alignment) and $\beta \gg \theta_E$ (large angular separation between source and lens), and calculate the difference $(\theta-\beta)$ in these regimes.

For the primary image $\theta_+$, we have:
\begin{equation}
\mbox{a)} \; \beta \ll \theta_E, \quad \theta_+ - \beta \simeq \theta_E \propto \sqrt{GM} \, ;
\end{equation}
\begin{equation}
\mbox{b)} \;  \beta \gg \theta_E, \quad \theta_+ - \beta \simeq \frac{\theta_E^2}{\beta} \propto \frac{GM}{b_0} \, .
\end{equation}
We see that the difference between impact parameters of lensed and unlensed rays ($b-b_0$) is large in case of good alignment.

For the secondary image $\theta_-$, we have:
\begin{equation}
\mbox{a)} \; \beta \ll \theta_E, \quad \theta_- - \beta \simeq - \theta_E   \propto \sqrt{GM}  \, ;
\end{equation}
\begin{equation}
\mbox{b)} \;  \beta \gg \theta_E, \quad \theta_- - \beta \simeq - \beta \, .
\end{equation}
We see that for secondary image the difference ($b-b_0$) is always significant. When $\beta \ll \theta_E$, it is proportional to $\sqrt{GM}$. When $\beta \gg \theta_E$, the difference is of the order of $b_0$: we may write $b-b_0 \simeq -D_d \beta = - b_0$.

\section{Limiting regime of nearly perfect alignment}
\label{sec:nearly-perfect}

The time delay in gravitational lensing is the difference between the travel time of a curved light ray for one of images and the travel time of an unlensed straight light ray from the source.
For the point-mass lens, the time delay can be written as (see, for example, \cite{Dodelson-2018})
\begin{equation} \label{td-def-2}
\Delta t(\theta) = \frac{1}{c} \frac{D_d D_s}{D_{ds}}\left[    \frac{1}{2} ( \theta - \beta )^2  - \theta_E^2 \ln {|\theta| }  \right]  ,
\end{equation}
or, with help of (\ref{le-point}), as a function of $\theta$ only,
\begin{equation} \label{td-def-3}
\Delta t(\theta) = \frac{1}{c} \frac{D_d D_s}{D_{ds}}\left[ \frac{1}{2} \left( \frac{\theta_E^2}{\theta} \right)^2  - \theta_E^2 \ln {|\theta|}  \right]  .
\end{equation}
Time delay has the same expression for primary and secondary images, only different $\theta$ should be substituted ($\theta_+$ or $\theta_-$).
The time delay expression also contains an additive constant with respect to $\theta$ which is usually not written. In practice this constant is usually not important because in observations of strong lens systems we always compare time delays from different images between each other.

For further convenience, let us define the two time delay terms, Shapiro (potential) delay
\begin{equation} \label{pot-point}
\Delta t^{pot}(\theta) \equiv - \tau_0 \, \theta_E^2 \ln {|\theta|}  \,  ,
\end{equation}
and geometrical delay
\begin{equation} \label{geom-point}
\Delta t^{geom}(\theta) \equiv \tau_0   \, \frac{1}{2} \left( \frac{\theta_E^2}{\theta} \right)^2   \,   .
\end{equation}
Here
\begin{equation}
\tau_0 = \frac{1}{c} \frac{D_d D_s}{D_{ds}} \,	.
\end{equation}
is a factor which only depends on the distance of the lens and the source, and does not affect the ratio between the geometric delay and potential delay. It is similar to the time delay distance (e.g.\cite{suyu2018}), but since here we do not consider the expansion of the universe, the extra dependence $(1+z_d)$ is neglected.

Formula (\ref{td-def-3}) is valid in  approximations used in lens equation (\ref{le}) itself. In particular, all angles $\hat{\alpha}$, $\beta$, $\theta$ should be small ($\ll 1$). Since $\theta$ is very small, the potential delay (\ref{pot-point}) is always positive.

When the angular separation between the image and the lens becomes larger, both terms in (\ref{td-def-3}) become smaller.
In other words, if $\theta_2 > \theta_1$, then $\Delta t^{pot}(\theta_2) < \Delta t^{pot}(\theta_1)$ and $\Delta t^{geom}(\theta_2) < \Delta t^{geom}(\theta_1)$. Both terms in time delay formula are positive.\\

Let us consider the situation when the angular position of the source is small compared to the Einstein ring size,
\begin{equation}
\beta \ll \theta_E \, ,
\end{equation}
see Fig.~\ref{fig:two-regimes}a. With such a condition, the expressions (\ref{sols-point}) reduce to
\begin{equation}
\theta_\pm \approx \pm \theta_E + \frac{\beta}{2} \, , \quad \theta_+ \gtrsim \theta_E \, , \quad |\theta_-| \lesssim \theta_E  \,  .
\end{equation}
In this case, the positions of the two lensed images are close to the Einstein ring: $|\theta| \simeq \theta_E$.
The two time delay terms $(\ref{pot-point})$ and $(\ref{geom-point})$ become
\begin{equation} \label{both-terms-nearly-1}
\Delta t^{pot}_\pm \simeq \tau_0 \, \theta_E^2 \left( \ln \frac{1}{|\theta_E|} \mp \frac{\beta}{2 \theta_E} \right) \, ,
\end{equation}
\begin{equation} \label{both-terms-nearly-2}
\Delta t^{geom}_\pm \simeq \frac{1}{2} \, \tau_0 \, \theta_E^2 \left(  1 \mp \frac{\beta}{\theta_E} \right)  \,  .
\end{equation}
Interestingly, linear corrections ($\beta/\theta_E$) are the same for potential and geometrical terms.

Since $\theta_E \propto \sqrt{GM}$, both delays have the same order of $GM$. We conclude that in this limiting regime of lensing the geometrical term is of the first order and should be taken into account together with potential term.

The condition $\beta \ll \theta_E$ or, equivalently, $\theta \sim \theta_E$ can be considered as the approximate condition required for strong lensing. In strong lens systems there is usually $\theta \simeq \theta_E$, and multiple images are formed with approximately the same order of magnification.
We conclude that the geometrical delay should be taken into account for every image in case of strong lens systems.

This conclusion agrees with the derivation of the time delay in \cite{Keeton-2018} and \cite{Keeton-2005}, discussed above in Sec.\ref{sec:shap-deriv}. In the literature, authors assume from the outset of the derivation that $\theta \sim \theta_E$ to obtain the time delay expression in the form used in gravitational lensing. This corresponds to our limiting regime of nearly perfect alignment. See also discussions on p.451 of \cite{Weinberg-2008} and p.269 of \cite{Sigl-2017}.

Additionally, our results agree with the paper \cite{Hackmann-2019}, where the authors calculate the exact time delay from a pulsar in Schwarzschild space-time numerically and compare it with different approximate formulas. It is claimed that the geometric delay is most significant when the emitting object is directly behind the black hole. Such a configuration of source, black hole and observer corresponds to our regime of nearly perfect alignment.

In order to complete our analysis, we calculate the difference between the primary and secondary images in the case of $\beta \ll \theta_E$. We obtain:
\begin{equation}
\Delta t^{geom} =  \Delta t^{geom}_+ - \Delta t^{geom}_- \simeq - \, \tau_0 \, \theta_E \, \beta  \,  ,
\end{equation}
\begin{equation}
\Delta t^{pot} =   \Delta t^{pot}_+ - \Delta t^{pot}_-   \simeq - \, \tau_0 \, \theta_E \, \beta .
\end{equation}
It is interesting to see that the potential and geometrical terms give the same contribution to the total difference $\Delta$ between the two images:
\begin{equation}
\Delta = \Delta t_+ - \Delta t_-  \simeq - 2 \, \tau_0 \, \theta_E \, \beta .
\end{equation}
This agrees with the formulae from Krauss and Small \cite{Krauss-Small-1991} where the time delay between two images was considered. They have stated that in the limit of small $\beta$ (in our notation) both terms in the time delay between two images become identical.

\section{Limiting regime of large angular separation between the source and the lens}
\label{sec:large-ang-sep}

Let us consider a source located far from the lens, i.e.
\begin{equation}
\beta \gg \theta_E  \,  ,
\label{eq:big-beta}
\end{equation}
see Fig.~\ref{fig:two-regimes}b. Note that under this condition, $\beta$ cannot be arbitrarily large, because both $\beta$ and $\theta_E$ are restricted to be $\ll 1$.
From expression (\ref{sols-point}) we obtain the angular positions of the primary and secondary images:
\begin{equation}
\theta_+ \approx \beta + \frac{\theta_E^2}{\beta} \, , \quad \theta_- \approx - \frac{\theta_E^2}{\beta}   \,  .
\end{equation}
The primary image is close to the unlensed source whereas the secondary image is close to the lens opposite the source.

The potential and geometrical time delay terms for the primary image are:
\begin{equation} \label{pot-s}
\Delta t^{pot}_+ = \tau_0 \, \theta_E^2 \ln \frac{1}{|\theta_+|} \simeq  \tau_0 \, \theta_E^2 \ln \frac{1}{|\beta|}  \,  ,
\end{equation}
\begin{equation} \label{geom-s}
\Delta t^{geom}_+ = \tau_0   \, \frac{1}{2} \left( \frac{\theta_E^2}{\theta_+} \right)^2  \simeq
\tau_0   \, \frac{1}{2} \left( \frac{\theta_E^2}{\beta} \right)^2   \,  .
\end{equation}

Since $\theta_E \propto \sqrt{GM}$, in this limit the potential term (\ref{pot-s}) is of the first order in ($GM$), whereas the geometrical term (\ref{geom-s}) is of the second order. We conclude that for the primary image we can neglect the geometrical term in comparison with potential one,
\begin{equation}
\Delta t^{geom}_+ \ll \Delta t^{pot}_+ \, ,
\end{equation}
which agrees with Shapiro's relationship. Moreover, we see that in this limit the Shapiro formula (\ref{Shap-b}) is recovered. Indeed, the potential term (\ref{pot-point}) contains $\theta$, so it depends on $b$ but not on $b_0$. But in the limit $\beta \gg \theta_E$ we obtain $\beta$ instead of $\theta$ in (\ref{pot-s}), and with $\beta=b_0/D_d$ we recover (up to an additive constant) the formula (\ref{Shap-b}). In more detail:
\begin{equation} \label{pot-s-expl}
\Delta t^{pot}_+  \simeq  \tau_0 \, \theta_E^2 \ln \frac{1}{|\beta|}  = \frac{1}{c} \frac{D_d D_s}{D_{ds}} \, \frac{4GM D_{ds}}{c^2 D_d D_s} \ln \frac{D_d}{b_0} = 
\end{equation}
\[
\quad \quad \:\:\, = \frac{4GM}{c^3} \ln \frac{D_d}{b_0} = \frac{2GM}{c^3} \ln \frac{D_d^2}{b_0^2} = \frac{2GM}{c^3} \ln \frac{4 D_d^2 D_{ds}}{4 D_{ds} b_0^2}  =
\]
\[
\quad \quad \:\:\, = \frac{2GM}{c^3} \ln \frac{4 D_d D_{ds}}{b_0^2} + \frac{2GM}{c^3} \ln \frac{D_d}{4 D_{ds}} \, .
\]
Here the first term coincides with (\ref{Shap-b}), whereas the second term is a constant that does not depend on $b_0$.

Solar system measurements of time delay occur in this limiting regime of lensing. Indeed, since the 'sources' in the Shapiro experiment are within the solar system, i.e. planets or satellites, all distances $D_d$, $D_{ds}$ and $D_s$ are of the same order. We can estimate the Einstein angular radius (\ref{einst-radius}) as
\begin{equation} \label{einst-solar-01}
\theta_E = \sqrt{ \frac{4G M_\odot D_{ds}}{c^2 D_d D_s} } \approx \sqrt{ \frac{4G M_\odot}{c^2 D} }  \simeq 41 \; \mbox{arcsec}   \,  ,
\end{equation}
where we have substituted $D \simeq D_d = 1$ AU (distance from Earth to Sun). Whereas a magnitude of $\beta$ can be estimated as the angular radius of Sun as seen from the Earth:
\begin{equation}
\beta \sim \theta_\odot = \frac{R_\odot}{D_d} \simeq 960 \; \mbox{arcsec} \, .
\end{equation}

We conclude that in time delay experiments in the Solar system we deal with the regime $\beta \gg \theta_E$ and the primary image only. Therefore, the geometrical delay can be neglected, which agrees with Shapiro's statement.\\

Let us now consider the secondary image. We should note that with the condition $\beta \gg \theta_E$, the secondary image usually is faint and difficult to observe. Therefore, for practical circumstances the time delay of the secondary image is likely unobservable.

For the secondary image, the two time delay terms are
\begin{equation}
\Delta t^{pot}_- = \tau_0 \, \theta_E^2 \ln \frac{1}{|\theta_-|} \simeq \tau_0 \, \theta_E^2 \ln \frac{1}{\theta_E^2} + \tau_0 \, \theta_E^2 \ln |\beta|   \,  ,
\end{equation}
\begin{equation}
\Delta t^{geom}_- = \tau_0   \, \frac{1}{2} \left( \frac{\theta_E^2}{\theta_-} \right)^2  \simeq
\tau_0   \, \frac{1}{2} \beta^2.
\end{equation}
With $\beta \gg \theta_E$, we can easily see that
\begin{equation} \label{eq:4-1-10}
\Delta t^{geom}_- \gg \Delta t^{pot}_- \, .
\end{equation}
We conclude that for the secondary image, the behaviour is reversed compared with the primary image: the geometrical term is much bigger than the potential term.

Together with the results of the previous Section, we may formulate the following statement: the geometrical delay should be taken into account as long as the image position $\theta$ is about $\theta_E$ or smaller. This statement is applicable for primary image in case of $\beta \ll \theta_E$ and for secondary image in both regimes. Another consequence from our discussion is the following: the geometrical delay always should be taken into account for the secondary image.

Additionally, it is possible to see that
\begin{equation}\label{eq:4-1-11}
\Delta t^{geom}_+ \ll \Delta t^{geom}_- \, , \quad \Delta t^{pot}_+ < \Delta t^{pot}_-   \, .
\end{equation}
In observations of strong lensing, we can only measure the total difference between the two images:
\begin{equation}
\Delta = \Delta t_+ - \Delta t_- = \Delta t^{geom}_+ + \Delta t^{pot}_+  -  \Delta t^{geom}_- -   \Delta t^{pot}_-
\end{equation}
Using conditions (\ref{eq:4-1-10}) and (\ref{eq:4-1-11}), we can say that the difference is determined mostly by the geometrical delay of the secondary image:
\begin{equation}
\Delta \simeq  -  \Delta t^{geom}_-.
\end{equation}
This conclusion agrees with Krauss and Small \cite{Krauss-Small-1991} where the time delay between two images was considered. They have stated that in the limit of large $\beta$ the geometric term dominates.

\section{Summary}
\label{sec:summary}

(i) In this paper we have discussed the contribution of the geometrical term (the term related to the  change of path length) to time delay of light from a distant source when a ray propagates to an observer near a gravitating body (lens). Our discussion is inspired by the following apparent discrepancy: in the derivation of the Shapiro delay, the change in path length due to light bending is considered as a negligible effect at second order. At the same time, the  geometrical delay is routinely taken into account in gravitational lensing together with the potential (Shapiro) term.

(ii) To explain the difference between these two situations, we have discussed the difference between the impact parameter $b$ of a lensed ray and the impact parameter $b_0$ of an unlensed straight line ray, in the situation when both of these rays have the same origin and end points. The Shapiro delay is derived with the approximation $b \simeq b_0$.
We have shown that such an approximation cannot be applied for the case when the angular position of the source is close to the observer-lens line. Restrictions on the validity of the approximations used during standard derivations of the Shapiro delay that arise in the case of good alignment are discussed in detail in Sec.\ref{sec:shap-deriv}.

On the other hand, in gravitational lens theory it is always assumed that $b \neq b_0$. In typical situations considered in strong lensing the difference between impact parameters of the lensed and unlensed rays is considerable and cannot be neglected. Therefore the geometrical change of path and correspondingly the geometrical delay connected with the change of path length should both be taken into account. The geometrical delay becomes crucially important when there is a good alignment of source, lens and observer and multiple images are formed.

(iii) To describe the problem quantitatively, we have introduced two regimes of lensing: nearly perfect alignment, $\beta \ll \theta_E$, and large separation, $\beta \gg \theta_E$, see Fig.~\ref{fig:two-regimes}. For these two cases we have investigated the behaviour of $(b-b_0)$ depending on $\beta$. Then, we have used the standard time delay expression from gravitational lens theory consisting of potential and geometrical delay, and have obtained the following properties for the primary image:

a) $\beta \ll \theta_E$. In this case, $b_0 \to 0$ and the difference between $b$ and $b_0$ becomes significant. The deflection angle is proportional to Einstein ring angular radius, $\hat{\alpha} \propto \theta_E \propto \sqrt{GM}$, and the geometrical term is of the same order ($GM$) as the potential (Shapiro) term. In this case the geometrical term in time delay cannot be neglected.

b) $\beta \gg \theta_E$. In this case, the difference between $b$ and $b_0$ for the primary image is negligible, the deflection angle $\hat{\alpha} \propto GM$, and the geometrical term in the time delay can be neglected because it is proportional to $\propto (GM)^2$. In this regime, the Shapiro formula for the delay is recovered, and the geometrical effect is of the second order, as stated by Shapiro.

(iv) In accordance with these two regimes, we have suggested the following explanation of the apparent contradiction to which this work is devoted.

In strong lens systems with multiple images we deal with the case of good alignment ($\beta \ll \theta_E$, or, at least, $\beta \lesssim \theta_E$). Therefore it is necessary to take into account the geometrical part in the calculation of the time delay for every image.

The original paper of Shapiro presents formulas for experiments with light propagation near the Sun. In such cases we deal with the primary image only and in the regime $\beta \gg \theta_E$. Therefore, the geometrical contribution can indeed be neglected, as stated by Shapiro.

(v) Regarding the secondary image, the geometrical term should be taken into account in both regimes of lensing (however, note that for $\beta \gg \theta_E$ the secondary image is hardly observable). Thus, we can formulate the following rule valid for both primary and secondary images: the geometrical delay should be taken into account as long as the image position $\theta$ is about $\theta_E$ or smaller. In particular, it means that the geometrical delay should be always be taken into account when we calculate the time delay difference between primary and secondary images.

\section*{Acknowledgements}

We thank the Referees for important comments.

The work of OYuT and GSBK was partially supported by the Russian Foundation for Basic Research and Deutsche Forschungsgemeinschaft according to the research project No. 20-52-12053. OYuT is thankful to Volker Perlick for many useful and pleasant
discussions throughout the work on this paper.

OYuT is grateful to DAAD (Deutscher Akademischer Austauschdienst) for support of his visit to ZARM (Zentrum f{\"u}r Angewandte Raumfahrttechnologie und Mikrogravitation), Bremen University, from 15 December 2018 until 31 January 2019, where a part of this work was discussed. OYuT expresses his gratitude to Prof C.~L{\"a}mmerzahl and his group for warm hospitality during this visit.
OYuT would like to thank the SWIFAR visiting fellow
program for support of his visit to South-Western Institute for Astronomy Research (SWIFAR) at Yunnan
University, from 21 April 2019 until 10 May 2019, where a part of this work was discussed.
OYuT expresses his gratitude to Prof Xiaowei Liu for invitation and kind hospitality.

X.E. is supported by NSFC grant No. 11873006.

\section*{Appendix: Angular deflections in two different lensing regimes}

The difference between the two regimes of lensing can be illustrated not only by the time delay components but also using an example of light deflection itself.

For light deflection, we can also formulate an apparent discrepancy of similar origin as in case of time delay. In the case of a solar eclipse, it is stated that the change of stellar positions is actually $\hat{\alpha}$, so this change is proportional to $GM$. While in strong lens systems, the typical angular separation between images is of order of $\theta_E$, so it is proportional to $\sqrt{GM}$. This discrepancy can be also explained by introducing two specific regimes of lensing.

During a solar eclipse, one can consider the change of positions of the stars from the point of view of gravitational lensing. We measure the difference between position of a lensed star and position of the unlensed star. In the notation of lensing, it is $(\theta-\beta)$. In the considered case, we have $D_d \ll D_s \approx D_{ds}$.
Therefore, we obtain from the lens equation (\ref{le}) that
\begin{equation} \label{sun-03}
\theta - \beta = \frac{D_{ds}}{D_s} \hat{\alpha} \simeq \hat{\alpha}  \, .
\end{equation}
Therefore, during a solar eclipse, change of angular position is equal to the deflection angle.

In approximation that $D_d \ll D_s \approx D_{ds}$, we have the Einstein radius equal to (compare with eq.(\ref{einst-solar-01}))
\begin{equation} \label{einst-solor-02}
\theta_E = \sqrt{ \frac{4GM D_{ds}}{c^2 D_d D_s} } \approx \sqrt{ \frac{4GM}{c^2 D_d} }  \simeq 41 \; \mbox{arcsec}   \,  .
\end{equation}

The angular radius of the Sun ($\theta_\odot$) is large in comparison with its Einstein angular radius. Thus during solar lensing, we are dealing with the case of $\beta \sim \theta_\odot \gg \theta_E$. In this limiting regime of lensing, we have for the primary image:
\begin{equation}
\theta_+ - \beta \simeq  \frac{\theta_E^2}{\beta} \propto \frac{GM}{b_0} \, .
\end{equation}
We have shown that
solar lensing corresponds routinely to the case $\beta \gg \theta_E$. The change of angular position $(\theta - \beta)$ is of order $GM$.

At the same time, the standard scenario considered in gravitational lensing is the case of multiple imaging, which approximately corresponds to our another limiting regime of lensing, $\beta \ll \theta_E$. In this case the change of angular position for every image is of the order of Einstein radius:
\begin{equation}
\theta -\beta \simeq \theta_E \propto \sqrt{GM} \, .
\end{equation}

For the sake of interest, let us calculate the distance $D_d$ where the Einstein radius of the Sun equals to the angular size of the solar radius $\theta_\odot = R_\odot/D_d$. In approximation $D_d \ll D_{s} \approx D_{ds}$, let the Einstein radius equal to the angular size of the Sun:
\begin{equation}
\sqrt{ \frac{4GM_\odot}{c^2 D_d} } = \frac{R_\odot}{D_d}.
\end{equation}
The distance can be easily solved, and we obtain
\begin{equation}
D_d = \frac{c^2 R_\odot^2}{4GM_\odot} \simeq 5 \cdot 10^2 \; \mbox{AU} \, .
\end{equation}
At this distance the Einstein radius of the Sun will become visible, i.e. larger than the radius of the solar disk.

Gravitational lensing by the Sun was also discussed in series of recent works by Turyshev and Toth \cite{Turyshev-2017, Turyshev-Toth-2017, Turyshev-Toth-2019a, Turyshev-Toth-2019b, Turyshev-Toth-2020}, including wave effects and refraction due to the plasma of the solar atmosphere.

\section*{References}

\bibliographystyle{ieeetr}

\end{document}